\def \mrm{\mathrm}
\def \del{\partial}
\def \spi{\mrm{spi}}
\def \elspi{\ell_\mrm{spi}^\mrm{eq}}

\documentclass[preprint,showpacs,preprintnumbers,amsmath,amssymb]{revtex4}

\usepackage{graphicx}
\usepackage{dcolumn}
\usepackage{bm}
\usepackage[dvips]{color}

\usepackage{amssymb}

\begin{document}





\title{Static and dynamical aspects of the metastable states \\
of first order transition systems}


\author{Tomoaki Nogawa}
\email{nogawa@serow.t.u-tokyo.ac.jp}
\author{Nobuyasu Ito}
\affiliation{
Department of Applied Physics, 
The University of Tokyo, Hongo, Bunkyo-ku, Tokyo 113-8656, Japan \\
}
\author{Hiroshi Watanabe$^1$}
\affiliation{
The Institute for Solid State Physics, 
The University of Tokyo, Kashiwanoha 5-1-5, Kashiwa, Chiba 277-8581, Japan$^1$
}

\begin{abstract}
We numerically study the metastable states of the 2d Potts model.
Both of equilibrium and relaxation properties are investigated 
focusing on the finite size effect.
The former is investigated by finding the free energy extremal point by the Wang-Landau sampling 
and the latter is done by observing the Metropolis dynamics 
after sudden heating.
It is explicitly shown that with increasing system size 
the equilibrium spinodal temperature approaches 
the bistable temperature in a power-law 
and the size-dependence of the nucleation dynamics agrees with it.
In addition, we perform finite size scaling of the 
free energy landscape at the bistable point. 
\end{abstract}






\maketitle

\section{Introduction}
\label{}

Phase transition dynamics is ubiquitous in our everyday life. 
One of the most popular examples is boiling of water \cite{Watanabe10}. 
In spite of its familiarity, 
our quantitative understanding of such phenomena remains quite poor 
due to its strong nonequilibrium property. 
The most popular theory of these first order transition dynamics 
is the classical nucleation theory \cite{Gibbs28, Rikvold94}
based on the droplet excitation picture, 
which fundamentally consider the cost of surface energy 
and gain of bulk energy.
However it is beyond this theory 
how many nuclei are produced for unit volume 
in the macroscopic system \cite{Langer68}. 
It is necessary to consider multi-nucleation 
including the interaction between nuclei.

Although nucleation is essentially nonequilibrium process, 
equilibrium property is useful as a starting point. 
First order transition, however, 
leaves more unclear points even in equilibrium 
in comparison with second order transition. 
One reason is that the former is very hard to investigate 
with conventional numerical simulations such as Monte-Calro methods; 
in order to know fine metastable stricture 
we have to observe sufficiently large number of 
rare events, overcoming of free energy barrier. 
Extended ensemble methods, however, 
have made breakthrough to this problem \cite{Berg92, Hukushima96, Wang01}.  
In this article, we report the detailed analysis 
of equilibrium metastable states 
by the Wang-Landau sampling method 
and its relation to the transition dynamics.

\section{Equilibrium spinodal point}


To investigate the equilibrium metastability, 
we employ the Wang-Landau sampling method \cite{Wang01}, 
which yields the density (number) of states 
as a function of macroscopic extensive quantities 
such as internal energy. 
This method is quite efficient to sample 
metastable states which appears with low probability 
in the Boltzmann distribution 
since this method realize a flat energy histogram.
For 2d Potts model, 
we calculate density of states $g(E)$ as a function of the internal energy, 
$$
E = \sum_{<i,j> \in \mrm{n.n.}} \left( 1 - \delta_{\sigma_i \sigma_j} \right) 
\quad \mrm{with} \quad
\sigma = 0, 1, \cdots, q-1. 
$$
The Helmholtz's free energy $F(\beta)$ is written as 
$$
e^{-\beta F(\beta)} = \sum_X e^{-\beta E(X)} = \sum_E g(E) e^{-\beta E} 
\equiv \sum_E e^{ -\beta F(\beta;E) }
$$
where $\beta=1/k_B T$ is inverse temperature 
and $X$ denotes microscopic state. 
The extremal condition; 
$\del F(\beta;E)/\del E = 0$, 
gives the energy corresponding to the maximum or minimum probability 
in canonical ensemble at given $\beta$. 
It is rewritten as 
\begin{equation}
\beta = \frac{\del S(E)}{\del E} 
\qquad \mathrm{where} \qquad 
S(E) = \ln g(E).
\label{eq:beta-E}
\end{equation}
When $\beta$ is not monotonic function of $E$, 
there are multiple minima in canonical ensemble for given $\beta$ 
indicating metastability and hysteresis. 
The things may be more clear 
to regard $\beta$ in Eq.~(\ref{eq:beta-E}) 
as a response of thermodynamic function $S(E)$ 
against the change of $E$ in the microcanonical ensemble. 
In microcanonical ensemble, 
there are coexisting phase between two distinct pure phases. 
The phase transition between pure phase and coexisting phase 
has direct connection to the evaporation/condensation transition 
investigated in continuum and lattice gas system 
\cite{Biskup02, Neuhaus03, Binder03, Nussbaumer06}.

The q-state Potts model on square lattice takes a first order transition 
between the paramagnetic and ferromagnetic phases
for $q>4$ at $\beta_c = \ln( 1+\sqrt{q} )$. 
We investigate the system with $q=8$ but 
believe that qualitative behavior does not depend on $q$. 
Periodic boundary condition is imposed 
for $L\times L$ square lattice. 
We perform parallel computation to treat large size system. 
The energy region is divided into some parts \cite{Landau04} 
and each part is associated to individual thread. 
In order to enhance the relaxation and guarantee the ergodicity, 
we make overlap energy region for neighboring threads, 
where the exchange of spin configuration is allowed 
satisfying a detailed balance condition, 
$$
\frac{ W(\{X,Y\} \rightarrow \{Y,X\}) }
     { W(\{Y,X\} \rightarrow \{X,Y\}) }
=\frac{ g_i \left(E(Y) \right) g_j \left(E(X) \right) }
      { g_i \left(E(X) \right) g_j \left(E(Y) \right) }, 
$$
in the same spirit with the parallel tempering method\cite{Hukushima96}. 
Here, $g_i$/$g_j$ is a density of states calculated by $i$-th/$j$-th thread 
and $W(\{X,Y \} \rightarrow \{Y,X \})$ 
means the transition probability 
from the compound state, $X$ for the $i$-th thread 
and $Y$ for the $j$-th thread, to its exchanged state. 
The density of states is calculated for the system with $L=32-2048$ 
by using 128 threads at most.  

\begin{figure}[t]
\begin{center}
\includegraphics[trim=20 230 200 -225,scale=0.360,clip]{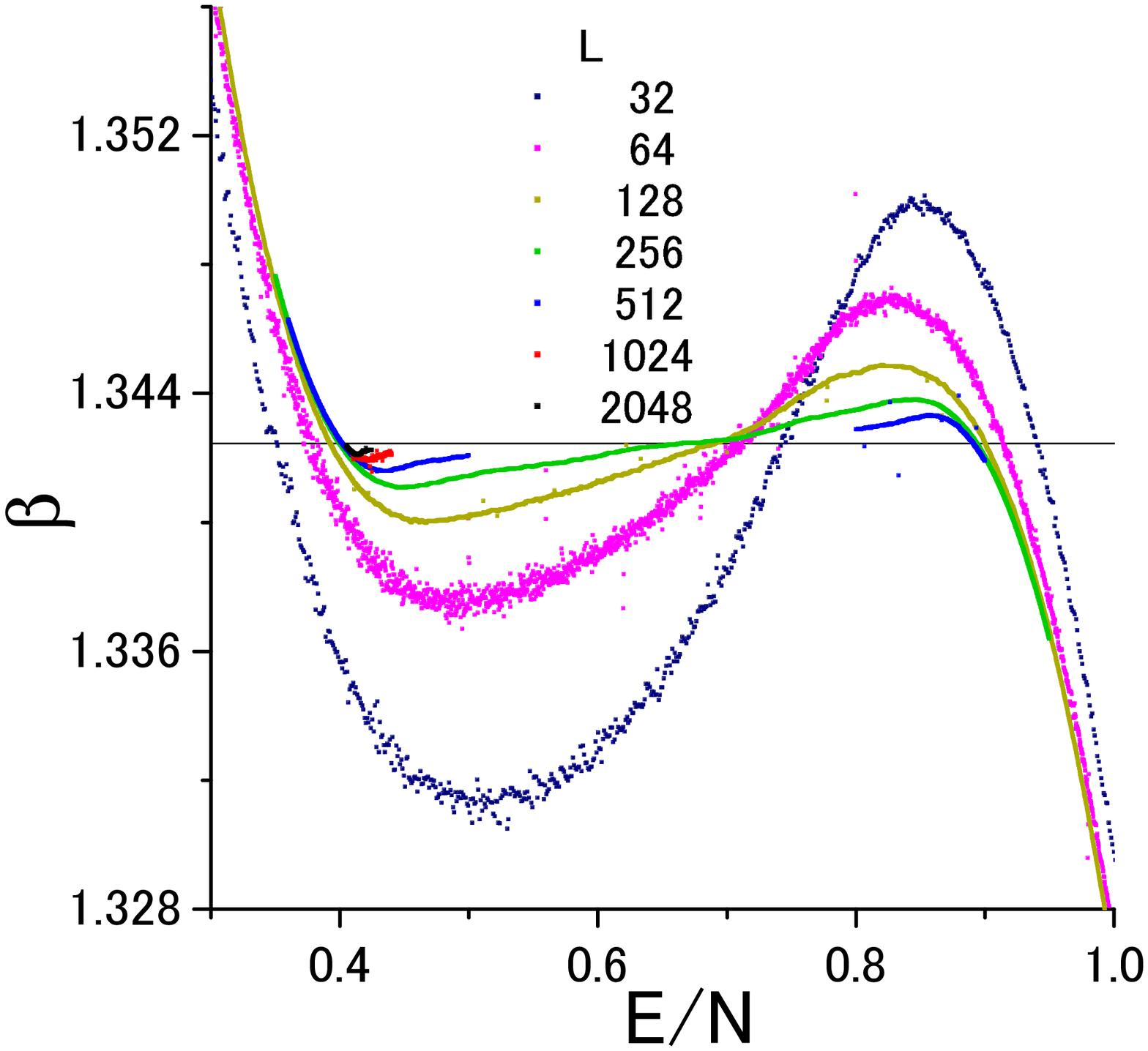}
 \hspace{0.5cm}
\includegraphics[trim=20 230 200 -225,scale=0.360,clip]{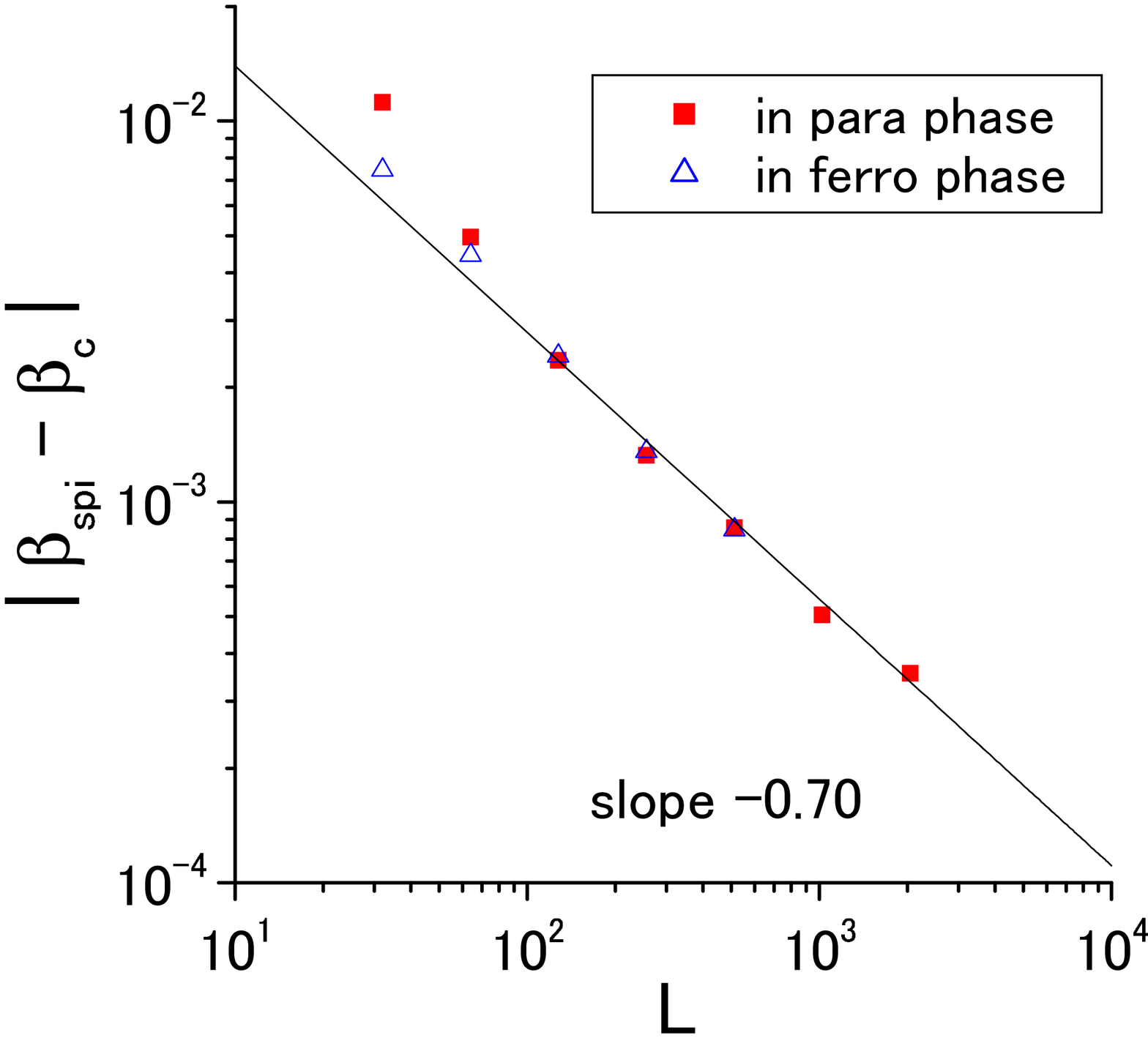}
\end{center}
\vspace{-5mm}
\caption{\label{fig:beta-E}
(left) 
inverse temperature as a function of internal energy 
calculated from difference, $\ln g(E+1) - \ln g(E)$, 
averaged over four individual simulations. 
The horizontal line indicates $\beta=\beta_c=\ln ( 1 +\sqrt{8} )$. 
(right) 
size-dependence of the deviation of spinodal temperature 
from the first order transition temperature. 
}
\end{figure}


The left panel of Fig.~\ref{fig:beta-E} shows relation 
between $\beta$ and $E$ with various system size. 
For given $\beta$ around $\beta_c=1.3424\cdots$,  
there are two free energy minimal points 
and one maximum one between them. 
There are two spinodal points, 
$(E_\spi^{f}, \beta_\spi^f)$ and 
$(E_\spi^{p}, \beta_\spi^p)$, 
at the end of low-energy ferromagnetic branch 
and high-energy paramagnetic branch 
as saddle node bifurcation points 
where energy minimal and maximal points meet to annihilate together. 
The both spinodal points explicitly depend on the system size 
and approach $\beta_c$ with increasing $L$, 
i.e., the hysteresis region, 
$\beta_\spi^f < \beta < \beta_\spi^p$, 
tends to disappear in the thermodynamic limit. 
The difference is plotted as a function of $L$ 
in right panel of Fig.~\ref{fig:beta-E}, 
which indicates 
\begin{equation}
| \beta_\spi - \beta_c |  \propto L^{-1/\nu}
\qquad \mrm{with} \qquad
\nu = 1.4(1), 
\label{eq:beta_spi}
\end{equation}
both for $\beta_\spi^f$ and $\beta_\spi^p$.

\begin{figure}[t]
\begin{center}
\includegraphics[trim=0 230 150 -225,scale=0.360,clip]{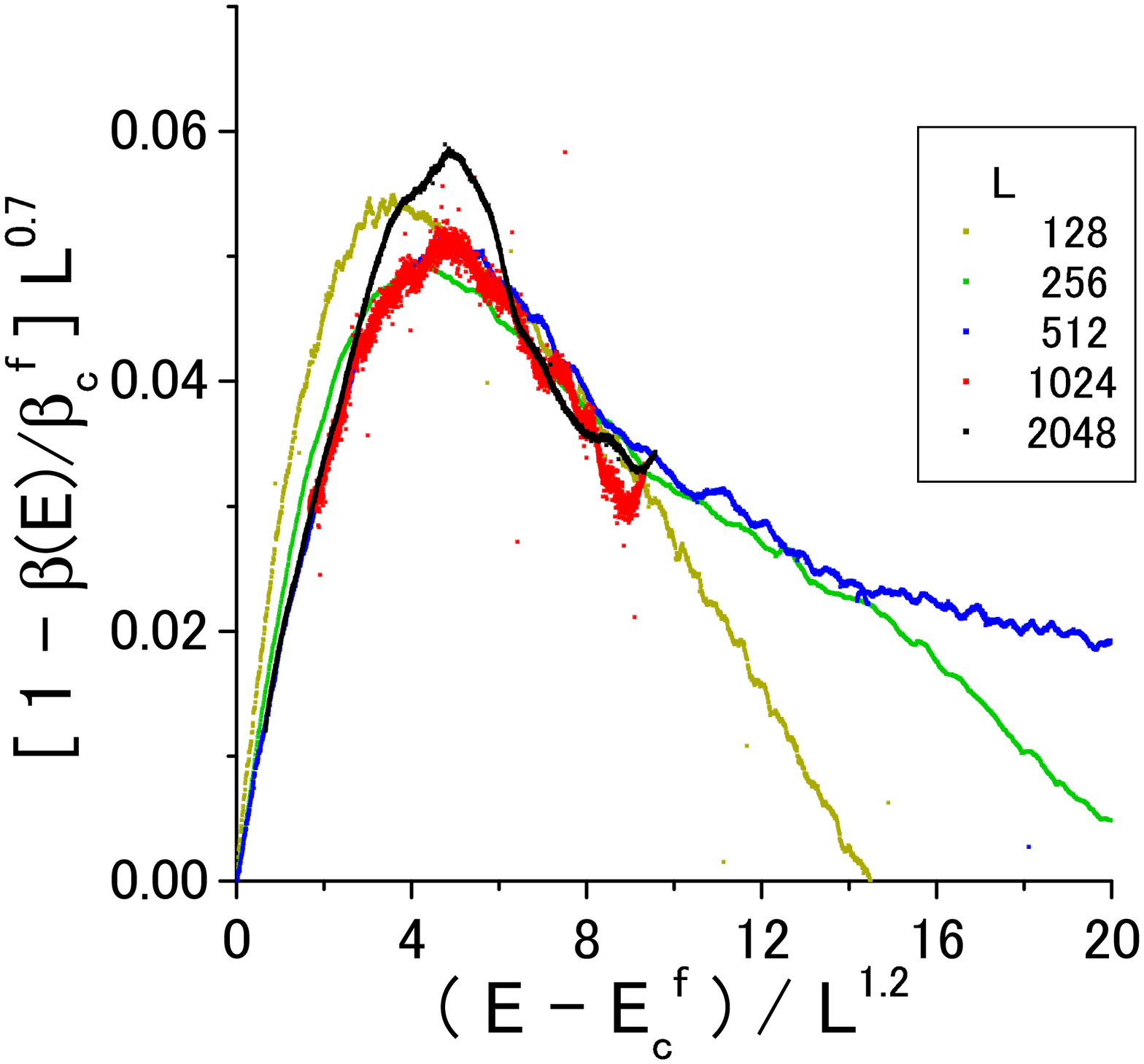}
\includegraphics[trim=20 230 150 -225,scale=0.360,clip]{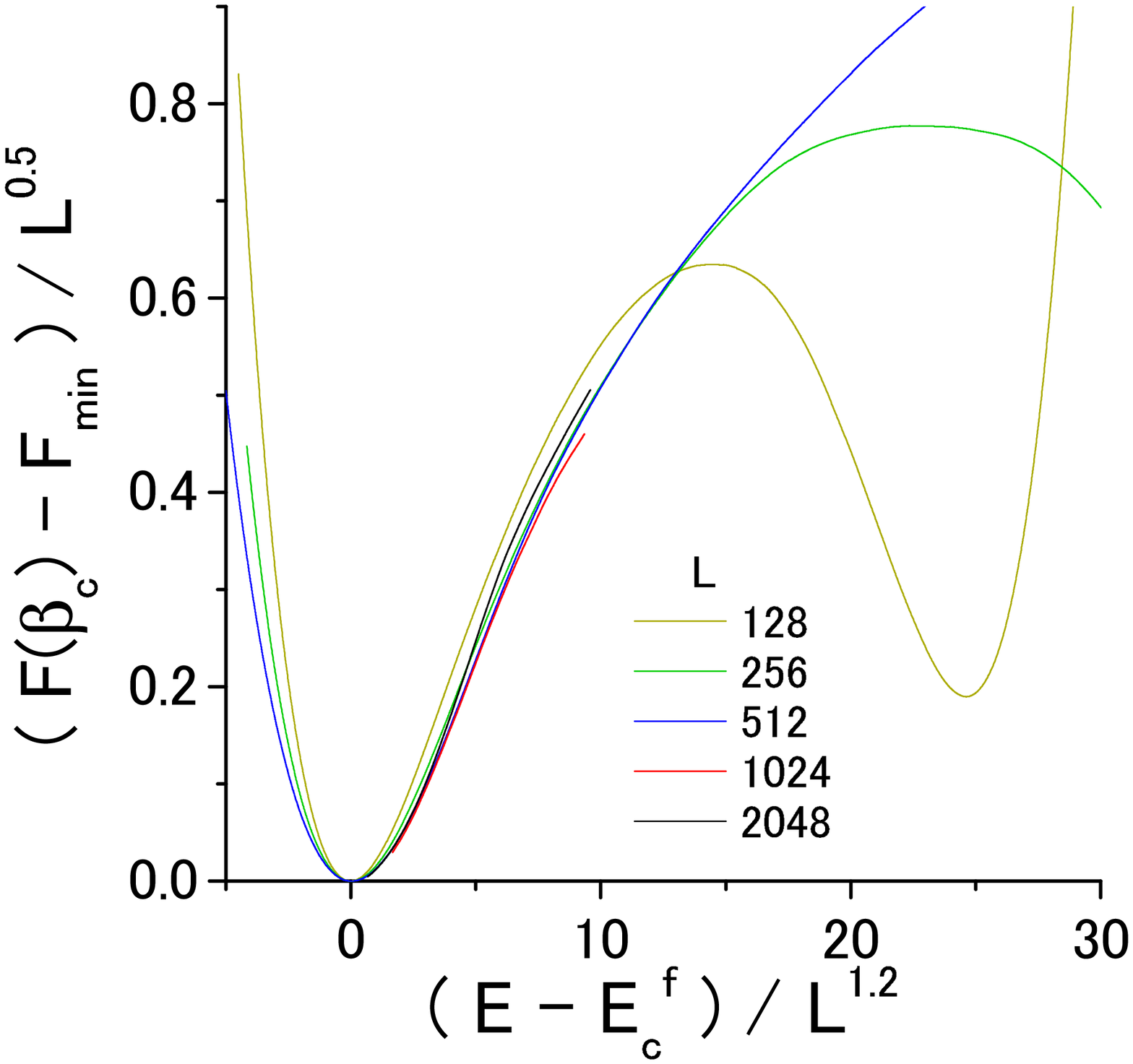}
\end{center}
\vspace{-5mm}
\caption{\label{fig:scaling}
finite size scaling of $\beta(E)$ (left) 
and $F(\beta_c,E)$ (right).
\label{fig:scaling}
}
\end{figure}

In addition, the spinodal energy $E_\spi^{f}/N$ and $E_\spi^{p}/N$ 
approaches $E_c^{f}/N \approx0.40$ and $E_c^{p}/N \approx 0.89$, 
respectively, with increasing $L$. 
These suggests a scaling behavior at the bistable point $\beta_c$; 
$$
| E_\spi - E_c |  \propto L^{d_E}
$$
and furthermore 
$$
F(\beta_c;E) = E - \beta_c^{-1} \ln g( E ) 
\approx F_\mrm{min} + L^{d_F} \hat{V} \left( (E - E_c^f)/L^{d_E} \right),
$$
for $E_c^f < E \ll E_c^p $. 
Hereafter we consider the scaling behavior around $E_c^f$ 
but the same argument holds around $E_c^p$. 
Scaled barrier function $\hat{V}(x)$ rises for $x \ll 1$ 
and flat for $x \gg 1$. 
This leads that 
$$
\beta(E) = \frac{\del \ln g(E)}{\del E} 
= \beta_c \left[ 
1 - L^{-(d_E-d_F)} \hat{V}' (E/L^{d_E}) \right]. 
$$
This reproduce Eq.~(\ref{eq:beta_spi}) with $d_E - d_F = 1/\nu$.
We perform scaling as shown in Fig.~\ref{fig:scaling} 
to estimate $d_E = 1.2(1)$ and $d_F = d_E - 1/\nu = 0.5(1)$. 
Good collapsing of data is obtained for $L \ge 256$. 
Note that this finite size scaling does not hold for $E<E_c^f$, 
where both internal energy and free energy are extensive, 
i.e., $d_E=d_F=1$. 
The free energy well becomes more asymmetric as system size increasing. 

Naive energetics of droplet excitation suggests $d_E=d-1=1$. 
The slight enhancement of the excitation energy is presumably 
due to the fractal roughness of the interface. 
On the other hand, 
the {\it dimension} of free energy barrier 
$d_F = 0.5(1)$ is smaller than 1.
This deviates from 
the behavior of macroscopic phase separation, $d_F=1$ \cite{Borgs92} 
and there should be another crossover for larger scale, 
$E-E_c^f \gg L^{0.5}$.

\section{Nonequilibrium relaxation, escape from metastable state}

\begin{figure}[t]
\begin{center}
\includegraphics[trim=20 230 150 -225,scale=0.360,clip]{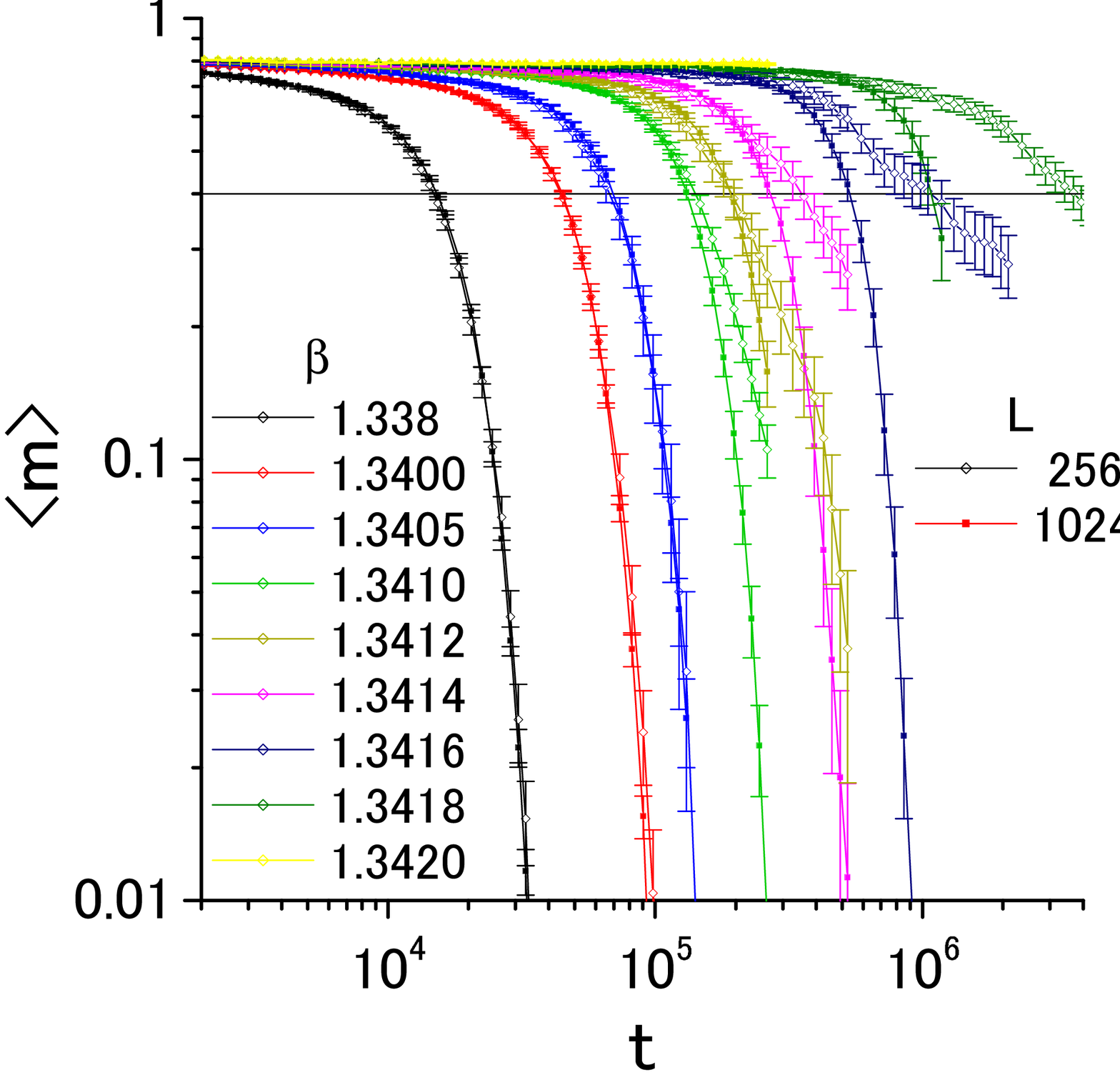}
\includegraphics[trim=20 230 150 -225,scale=0.360,clip]{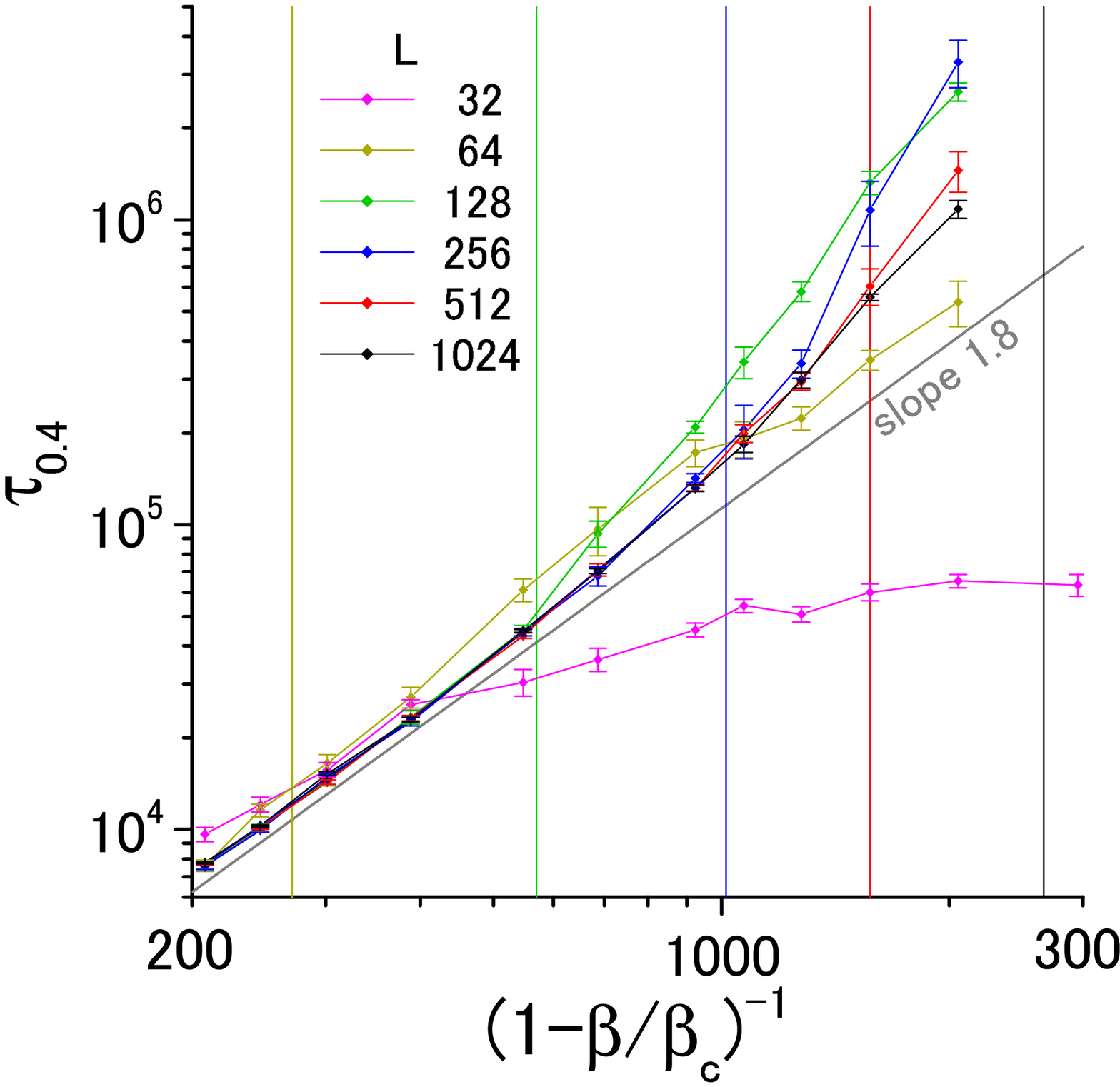}
\end{center}
\vspace{-5mm}
\caption{\label{fig:M-t}
(left) time evolution of order parameter for $L$=256 and 1024. 
The data are averaged over 16384/$L$ samples. 
(right) temperature dependence of the time crossing 
$m=0.4$ for various system size. 
The vertical lines denote the equilibrium spinodal temperatures 
for various system size. 
}
\end{figure}

Next, we investigate transition dynamics 
from ferromagnetic state to para state at fixed temperature. 
The time evolution of the system is given 
by the Metropolis algorithm \cite{Rikvold94, Krzakala11a}. 
The unit time is defined as the span 
in which all spins are updated one time in a sequential manner. 
All spins $\sigma_i$ are set to the same state named 0 
in the initial state at $t=-1000$.
After the system is equibrated at $\beta_c$,  
$\beta$ is suddenly lowered to aimed value at $t=0$. 

We observe an order parameter 
$\langle m \rangle = 
(N^{-1} \langle \sum_i \delta_{\sigma_i, 0} \rangle 
- q^{-1} )/(1-q^{-1})$, 
where $\langle \cdots \rangle$ means the average 
over independent random number realizations.  
Time evolution of $\langle m \rangle$ for $L$=256 and 1024 is shown 
in the left panel of Fig.~\ref{fig:M-t}. 
For $L$=1024, there is almost no finite size effect in this data. 
The characteristic decay time tends to diverge as approaching $\beta_c$. 
We plot the time $\tau_{0.4}$ at which $\langle m(t) \rangle$ crosses 0.40 
as a function of $\beta$ in right panel of Fig.~\ref{fig:M-t}. 
For sufficiently large $L$, size dependence is not observed 
and the divergence seems to be a little faster than a power-law type; 
$\tau \approx (1-\beta/\beta_c)^{-z\nu}$. 
This implies that the dynamics is of thermal activation type, 
typically $\tau \propto \exp[ \mrm{const.}/(\beta_c-\beta)]$, 
although it is difficult to speculate its analytic expression.  

On the other hand, finite size effect appears approaching $\beta_c$ 
the earlier, the smaller $L$ becomes. 
The inverse temperature above which 
$\tau_{0.4}$ deviates from large size behavior 
agrees well with the finite-size spinodal point $\beta_\spi^f(L)$ 
obtained by equilibrium simulations.
Initially $\tau_{0.4}$ overcomes the large size behavior, 
which presumably due to the backward transition 
from para phase to ferro phase, 
while $\tau_{0.4}$ becomes smaller than large size behavior 
and saturates to a certain value increasing with $L$.

\section{Summary}

In conclusion, 
we calculate the size-dependent spinodal points of the 8-state Potts model 
and perform finite size scaling to obtain 
fractal exponent for free energy barrier $d_F \approx 0.5$ 
and internal energy $d_E \approx 1.2$. 
These exponents can be compared with 
$(d^2-d)/(d+1)=2/3$ and $d^2/(d+1)=4/3$ in Ref.~\cite{Binder03}, 
where evaporation/condensation transition is discussed 
based on the Ising model in micromagnetic ensemble, 
by relating $\beta_\spi - \beta_c$ and $(E-E_c)/N$ 
to magnetic field $\tilde{H}_t^{(2)}$ 
and magnetization $m_t - m_\mrm{coex}$, respectively. 

Additionally it is confirmed that size dependent spinodal temperature 
also gives the dynamical criterion in transition dynamics, 
which suggests some connection between 
equilibrium metastability and the nonequilibrium relaxation. 
Equation~\ref{eq:beta_spi} implies that there is 
a diverging length scale in bulk 
$$
\elspi(\beta) \propto | \beta - \beta_c |^{-\nu}, 
$$
with $\nu \approx 1.4$. 
It is important to note that such critical like behavior 
is observed in the energy region of coexisting phase, 
which is not realized in the canonical ensemble. 
It will be very interesting if we can find any relation 
between these unphysical equilibrium state 
and the nonequilibrium transient state with corresponding energy. 
The exponent $\nu \approx 1.4$ is larger than 1 which is expected 
for critical nucleus radius 
and $\elspi$ may mean the average distance between critical nuclei. 
The advanced study is required for the thermally activated nuclear growth, 
where typical distance of nuclei obtained here will give a useful hint. 



\section*{Acknowledgement}

This work was partly supported by Award No. KUK-I1-005-04 made by King Abdullah University of Science and Technology (KAUST).












\end{document}